
\font\titlefont = cmr10 scaled \magstep2
\magnification=\magstep1
\vsize=22truecm
\voffset=1.75truecm
\hsize=15truecm
\hoffset=0.95truecm
\baselineskip=20pt

\settabs 18 \columns

\def\b{\bigskip}
\def\bb{\bigskip\bigskip}

\def\ce{\centerline}

\def\no{\noindent}

\def\ttbar{t\overline t}

\rightline{AMES-HET 94-11}
\rightline{October 1994}
\bb

\ce{\titlefont{Unitarity And Anomalous Top-Quark Yukawa Couplings}}

\b

\ce{ K. Whisnant, ~~Bing-Lin Young, and ~~X. Zhang}
\b
\ce{Department of Physics and Astronomy, }
\ce{ Iowa State University,}
\ce{ Ames, Iowa  50011}

\b
\bb
\ce{\bf ABSTRACT}
\b

\no ~~ Unitarity constraints on anomalous top-Higgs couplings are examined.
In the calculation, we considered all $t {\overline t}$ and two-boson channels
and obtained the maximal value of the coupled channel amplitude, which
sets the unitarity condition. We compare
the unitarity constraints with the constraints from
electroweak baryogenesis and electric dipole moments derived earlier. Tighter
constraints can be obtained by
including a $t {\overline t} H H$ contact term which arises from
one realization of our effective interaction.

\bb
\filbreak

Recently the CDF collaboration presented evidence for the top quark
with mass $m_t \sim 174$ GeV[1]. Since $m_t$ is of order the Fermi scale,
the top quark may hold the clue to the physics of the mass generation. In a
recent paper[2], we have studied the phenomenology of a
non-standard top quark Yukawa coupling.
In the notation of Ref.~2, the general interaction of the top
quark to the Higgs boson is given by
$${
{\cal L}^{eff}_t = {m_t \over v}~{\overline{t}}~~[
          ( 1 + \delta \cos\xi )~ + ~ i ~( \delta \sin\xi)~ \gamma_5 ] t ~ H
{}~~~, } \eqno(1)$$
\no where $\delta$ is a free parameter, $\xi$ is a CP phase, and both are
zero in the standard model. Such a non-standard top quark Yukawa coupling,
${\cal L}_t^{eff}$, is expected to appear in the dynamical electroweak
symmetry breaking theories, such as composite Higgs models[3] and (low scale)
top quark condensation models[4], and also in extensions of the standard model
with fundamental Higgs scalars. In Ref.~2 we sketched a derivation of
${\cal L}^{eff}_t$ in a class of the left-right symmetric models[5].
However, the essential motivation for the proposal of ${\cal L}^{eff}_t$
is based on the observation that during the
electroweak phase transition in the early universe, ${\cal L}_t^{eff}$
can help produce the baryon number asymmetry. In Ref.~2, we have shown that
for a CP violation quantity $\delta \sin\xi$ at the level of order $\sim 0.3$
we can understand the matter-antimatter asymmetry of the universe at the weak
scale. However, since the electroweak baryogenesis[6] calculations presently
available are qualitative and the quantitative results obtained so far are
only accurate to within a couple of orders of magnitude, direct measurements
of model parameters, e.g., at future colliders, will be needed to test the
various models of electroweak baryogenesis. In
Ref.[2], we have used electroweak baryogenesis and the electric dipole moments
of electron and neutron calculated using ${\cal L}_t^{eff}$ to place limits on
$\delta$ and $\xi$, and also discussed the possibility of testing
${\cal L}_t^{eff}$ directly in the next generation linear colliders. In this
paper we will examine the bounds on $\delta$ and
$\xi$ from partial-wave unitarity constraints.

In the standard model, there is a definite relationship
between the fermion mass and its coupling to the Higgs boson, which keeps
perturbative unitarity valid at all scales. With non-vanishing $\delta$
and $\xi$, unitarity will be violated at some scale ${\sqrt{s}}=\Lambda$, where
where one expects $\Lambda$ to be the approximate scale at which new
particles and interactions appear. The processes affected by the non-standard
$H \ttbar$ coupling involve the Higgs boson, top quark and weak gauge bosons.
First we calculate the amplitudes for $\ttbar\to\ttbar$ and $\ttbar\to
{\rm bosons}$. We will keep only terms of order $G_Fm_t\sqrt{s}$ or
$G_Fm_t^2$, whichever is larger for a given process. Terms of order
$G_Fm_W^2$, $G_Fm_bm_t$, $G_Fm_b^2$, or which vanish for large $s$ have
little effect on our results and are ignored. We will denote by $+$ or $-$
the helicities of the fermions (written in the order the fermions appear in
$\ttbar\to \ttbar$ or $\ttbar\to{\rm bosons}$). Since we will examine in
detail only the $J=0$ partial-wave amplitudes (which gives a stronger
constraint than those obtained from higher partial wave amplitudes), only the
$++$ and $--$ initial and final state helicity combinations will be discussed.
The external gauge bosons involved are longitudinal as they give the leading
high energy behavior in the scattering amplitudes.

For the process $\ttbar\to\ttbar$ the leading order helicity amplitudes are
all constant in $s$. For color-singlet initial and final states there are
$s$- and $t$-channel diagrams involving the photon, $Z$ boson and Higgs boson.
To leading order the color-singlet helicity amplitudes are
$$\eqalignno{
T_{++,++}(\ttbar\to\ttbar) &= T_{--,--}^*(\ttbar\to\ttbar)
= -3\sqrt2 G_F m_t^2 \left[1+|1+\delta e^{i\xi}|^2\right],
&(2a)\cr
T_{++,--}(\ttbar\to\ttbar) &= T_{--,++}^*(\ttbar\to\ttbar)
= 4\sqrt2 G_F m_t^2 \left[(1+\delta e^{i\xi})^2-1\right].
&(2b)\cr}$$
Note that the $T_{\pm\pm,\mp\mp}$ amplitudes in Eq.~2 vanish in
the standard model (where $\delta=0$). For color-octet states only the
$t$-channel diagrams contribute and the leading order helicity amplitudes are
$${\tilde T}_{++,--}(\ttbar\to\ttbar) = {\tilde T}^*_{--,++}(\ttbar\to\ttbar)
= \sqrt2 G_F m_t^2 \left[(1+\delta e^{i\xi})^2-1\right].
\eqno(2c)$$
In Eq.~2 helicity combinations with subleading amplitudes are not listed.

For the process $\ttbar\to W_L^+W_L^-$,
where $W_L$ is a longitudinal $W$ boson,
 there are $s$-channel diagrams involving
the photon, $Z$ boson and Higgs boson, and a $t$-channel diagram with a virtual
$b$-quark. In the standard model there are contributions to the amplitudes
which are proportional to $s$ and $\sqrt{s}$, but these cancel upon summing
over all diagrams. In our case with
an effective top-quark Yukawa coupling this cancellation is not assured, and
in fact the anomalous part of the Yukawa interaction gives an amplitude
proportional to $\sqrt{s}$ in helicity channels with a Higgs
boson intermediate state. The leading order color-singlet amplitudes are
$$T_{++}(\ttbar\to W_L^+W_L^-) = -T_{--}^*(\ttbar\to W_L^+W_L^-) =
\sqrt6 G_F m_t\sqrt{s}\left[\delta e^{i\xi}\right].
\eqno(3)$$
For $\ttbar\to Z_LZ_L$, there are diagrams with an $s$-channel Higgs boson and
a
$t$- and $u$-channel top quark; the leading order color-singlet amplitudes are
$$T_{++}(\ttbar\to Z_L Z_L) = -T_{--}^* (\ttbar\to Z_L Z_L)
= \sqrt6 G_F m_t\sqrt{s}\left[\delta e^{i\xi}\right].
\eqno(4)$$
For $\ttbar\to Z_LH$, there are diagrams with an $s$-channel $Z$ boson
and a $t$- and $u$-channel top quark; the leading order color-singlet
amplitudes are
$$T_{++}(\ttbar\to Z_L H) = T_{--}^* (\ttbar\to Z_L H)
= -\sqrt6 G_F m_t\sqrt{s}\left[\delta e^{i\xi}\right].
\eqno(5)$$
For $\ttbar\to HH$ the leading color-singlet contribution involving the
$t\overline tH$ vertex is from an $s$-channel Higgs
$$T_{++}(\ttbar\to HH) = -T_{--}^* (\ttbar\to HH)
= -3\sqrt6 G_F m_H^2 m_t\left[1+\delta e^{i\xi}\right]/\sqrt{s};
\eqno(6)$$
the diagram with a $t$- or $u$-channel top quark does not contribute to the
$J=0$ amplitude. For $m_H\approx m_Z$ (our region of interest
\footnote{[F.1]}{This is required by baryogenesis
to avoid washout of the asymmetry[8].}
) and $m_t=174$~GeV the amplitude
in Eq.~6 may be neglected, although for $m_H \geq \sqrt{m_t\sqrt{s}}$
it should be included.
Finally, the color-octet amplitudes vanish for $\ttbar\to{\rm bosons}$.

With the expressions for the amplitudes in Eqs.~2 to 6, we may now determine
the constraints from partial-wave unitarity. The $J=0$ partial-wave amplitude
for a process with amplitude $T$ is
$$a_0={1\over32\pi}\int_{-1}^1 T ~d(\cos\theta).
\eqno(7)$$
Partial-wave unitarity implies that $|a_0|<1$. The most restrictive bound
comes from the largest eigenvalue of the coupled channel $T$-matrix. If we
write the channels in the order $t_+\overline t_+$, $t_-\overline t_-$,
$W_L^+W_L^-$, $Z_LZ_L/\sqrt2$, $Z_LH$, and $HH/\sqrt2$, then the coupled
channel matrix for the color-singlet $J=0$ partial wave is
$$a_0={\sqrt2 G_F m_t^2\over16\pi} \pmatrix{
-T_1   &  T_2 &  T_3   &  T_3/\sqrt2   & -T_3   &  -T_4/\sqrt2\cr
 T_2^* & -T_1 & -T_3^* & -T_3^*/\sqrt2 & -T_3^* & T_4^*/\sqrt2\cr
 T_3^* & -T_3 &    0   &    0   &    0   &    0\cr
 T_3^*/\sqrt2 & -T_3/\sqrt2 &    0   &    0   &    0   &    0\cr
-T_3^* & -T_3 &    0   &    0   &    0   &    0\cr
 -T_4^*/\sqrt2 & T_4/\sqrt2 &    0   &    0   &    0   &    0\cr},
\eqno(8)$$
where
$$\eqalignno{
T_1 &= 3\left[1+|1+\delta e^{i\xi}|^2\right],
&(9a)\cr
T_2 &= 4\left[1+\delta e^{i\xi}\right]^2,
&(9b)\cr
T_3 &= \sqrt3(\sqrt{s}/m_t)\left[\delta e^{i\xi}\right],
&(9c)\cr
T_4& = 3 {\sqrt{3}} m_H^2/ (m_t {\sqrt{s}})\left[1+\delta e^{i\xi}\right]. &
 (9d) \cr
}$$
In Eq.~8 we must also include all channels of the type bosons$\to$bosons. The
$t$-quark Yukawa coupling does not contribute to these processes at the tree
level, so these amplitudes assume their standard model values, which are of
order $G_Fm_H^2$. Since in our scenario $m_H\approx m_Z$, this is small
compared to $G_F m_t \sqrt{s}$ and $G_Fm_t^2$, and we can set the purely
bosonic amplitudes to zero as indicated in Eq.~8. We have checked that the
amplitudes for these channels are at most a few percent of the dominant
amplitudes in the coupled-channel matrix for the values of the parameters we
are considering, which justifies this approximation. Similarly the
contribution of $T_4$ in Eq.~6 is small, and therefore the last row and column
in Eq.~8 can be neglected (for a different case with a non-negligible $T_4$,
see below).

The characteristic equation for the roots of the remaining $5\times5$ matrix
is easily found. After removing a trivial zero root, the root with the largest
magnitude may be found numerically. The corresponding upper bound on $\delta$
as a function of $\xi$ is shown as the dashed curve in Fig.~1, assuming
unitarity is preserved up to $\sqrt{s} = 1$~TeV. Also shown for comparison are
the approximate lower bound from electroweak baryogenesis and the upper bound
from the neutron electric dipole moment [2]. We should point out that the
constraints on $\delta$ and $\xi$ depend on the $\sqrt{s}$ at which we assume
that unitarity is saturated. We have also checked that if the magnitude of the
anomalous coupling is comparable to the standard model coupling, the scale of
the new physics should be no more than a few TeV[9].

Before concluding, we should also point out that a sizable $T_4$ can be
generated even for $m_H \simeq m_Z$ if higher dimension operators are
considered. Such operators can arise quite naturally in a linearly realized
$SU(2)_L \times U(1)_Y$ effective Lagrangian. For example, the operator[2]
$${
{\cal O}^t=\delta e^{i\xi}\left({\Phi^\dagger \Phi\over v^2}-{1\over2}\right)
{\overline{\pmatrix{t_L \cr b_L \cr}}}{\tilde\Phi} t_R + h.c.
{}~~} , \eqno(10)$$
where $\Phi$ is the complex standard model doublet scalar and
${\tilde\Phi} = i \sigma_2 \Phi^*$, will generate $t\overline tHH$ and
$t\overline tHHH$ contact terms in addition to the anomalous part of the
Yukawa coupling in Eq.~1. Since in the broken phase $\Phi\to(H+v)/\sqrt2$, the
ratio of the $t {\overline t}H H$ to $t {\overline t}H$ coupling strengths in
Eq.~10 is fixed to be $3/2v$. The contribution from the contact term
$t {\overline t} H H$ then dominates the color-singlet $t{\overline t}\to HH$
amplitude; we find the new amplitude
$$T_4=T_{++}(\ttbar\to HH) = -T_{--}^*(\ttbar\to HH) = 6T_3.
\eqno(11)$$
Now $T_4$ is proportional to $G_F m_t {\sqrt{s}}$ and the full $6\times6$
matrix in Eq.~8 must be used. The unitarity limit derived from the largest
eigenvalue in this case is shown as the solid curve in Fig.~1. We note that
in general for an effective lagrangian with
non-linear realization of $SU(2)_L \times U(1)_Y$ (where $H$ is considered a
scalar singlet) there is an independent free parameter associated with each
contact term, which makes it impossible to improve the unitarity bounds on
the anomalous Yukawa coupling by considering the effect of the contact terms
on the scattering amplitude. In this case the bound found by setting $T_4=0$
will be the upper limit.

In conclusion, we have examined the unitarity constraints on anomalous
top-Higgs couplings. Our results for $\Lambda=1$~TeV, where $\Lambda$ is
the mass scale of the new physics which produces such an anomalous top-Higgs
interaction, are summarized in Fig.1. From the figure we see that $\delta$,
the magnitude of the anomalous top Yukawa coupling, should lie in the range
0.3$-$3.5 in order to satisfy all of the constraints. One can also turn this
requirement around. We have checked that for $\delta \sin\xi \sim 0.3$, as
implied by electroweak baryogenesis[2], unitarity requires that $\Lambda$ is
less than about 5 TeV. Since this new physics scale is not too far from the
Fermi scale, one might expect that new physics effects will not only show
up as a modified Yukawa interaction (Eq.~1), but also in anomalous couplings
of the top quark to the gauge bosons, which will be examined later.

\b
\bb
This work is supported in part by the Office of High Energy and Nuclear
Physics of the U.S. Department of Energy (Grant No. DE-FG02-94ER40817).

\bb
\bb
\ce{\bf Figure Captions}

\b
\item{[Fig.1]} Bounds on a non-standard top-quark Yukawa coupling in terms of
the parameters $\delta$ and $\xi$. The solid curve is the upper bound from
unitarity using the $t\overline tH$ and $t\overline tHH$ anomalous interactions
and the dashed curve is the bound from the $t\overline tH$ interaction alone,
for $\Lambda=1$~TeV. The dotted curve is an approximate lower bound from
electroweak baryogenesis and the dash-dotted curve is an upper bound from the
experimental limit on the electric dipole moment of the neutron, taken from
Ref.~2.

\bb
\b

\vfill\eject

\ce{\bf References}
\b

\item{[1]}
CDF Collaboration, F. Abe et al, FERMILAB-PUB-94/097-E, 1994.

\item{[2]}X. Zhang,~ S. Lee,~ K. Whisnant and B.-L. Young, ``Phenomenology
of a non-standard top quark Yukawa coupling," Iowa State University Preprint,
AMES-HET-94-05, June 1994 (to appear in Phys. Rev. D).

\item{[3]}D.B. Kaplan and H. Georgi, Phys. Lett. B136, 183 (1984);
D.B. Kaplan, S. Dimopoulos and H. Georgi, Phys. Lett. B136, 187 (1984);
H. Georgi, D.B. Kaplan and P. Galison, Phys. Lett. B143, 152 (1985);
H. Georgi and D.B. Kaplan, Phys. Lett. B145, 216 (1984);
 M.J. Dugan, H. Georgi and D.B. Kaplan, Nucl. Phys. B254, 299 (1985);
V. Koulovassilopoulos and R.S. Chivukula, BUHEP-93-30 (Hep-Ph/9312317),
Dec. 20, 1993.

\item{[4]}C.T. Hill, Phys. Lett. B266, 419 (1991).

\item{[5]}J.C. Pati and A. Salam,
Phys. Rev. D10, 275 (1974); R.N. Mohapatra and J.C. Pati, {\it ibid.} D11,
566 (1975); D11, 2558 (1975); Senjanovic and R.N. Mohapatra,
{\it ibid.} D12, 1502 (1975).

\item{[6]}For a review, see, A. Cohen, D. Kaplan and A. Nelson,
          Ann. Rev. Nucl. Part. Sci. 43, 27 (1993); on recent development on
spontaneous baryogenesis, see, A.G. Cohen, D.B. Kaplan, A.E. Nelson, Phys.
Lett.
B336, 41 (1994), and references therein.

\item{[7]}For a discussion on unitarity bounds on the
Higgs-boson and top-quark masses, see, W. Marciano, G. Valencia, and
S. Willenbrock, Phys. Rev. D40, 1725 (1989), and references therein; for
a discussion on unitarity bounds on anomalous gauge boson couplings, see,
G. Gounaris, J. Layssac, J. Paschalis and
F. Renard, PM/94-28, THES-TP 94/08, August 1994.

\item{[8]}X. Zhang, B.-L. Young, Phys. Rev. D49, 563 (1994);
and references therein.

\vfill
\eject

\item{[9]}M. Golden, Harvard University Preprint,
HUTP-94/A020, July 1994, Hep-Ph/9408272.

\vfill
\eject
\bye